\documentclass[prl,superscriptaddress,twocolumn,preprintnumbers,
    showpacs,a4paper]{revtex4}

\usepackage{graphicx}% Include figure files
\usepackage{dcolumn}% Align table columns on decimal point
\usepackage{bm}% bold math

\begin{document}

\title{Activation measurement of the $^3$He($\alpha,\gamma$)$^7$Be cross section at low energy}

\author{D.~Bemmerer}\affiliation{Istituto Nazionale di Fisica Nucleare (INFN), Sezione di Padova, via Marzolo 8, 35131 Padova, Italy}
\author{F.~Confortola}\affiliation{Universit\`a di Genova and INFN Sezione di Genova, Genova, Italy}
\author{H.~Costantini}\affiliation{Universit\`a di Genova and INFN Sezione di Genova, Genova, Italy}
\author{A.~Formicola}\affiliation{INFN, Laboratori Nazionali del Gran Sasso (LNGS), Assergi (AQ), Italy}
\author{Gy.~Gy\"urky}\affiliation{Institute of Nuclear Research (ATOMKI), Debrecen, Hungary}
\author{R.~Bonetti}\affiliation{Istituto di Fisica Generale Applicata, Universit\`a di Milano and INFN Sezione di Milano, Italy}
\author{C.~Broggini}
 \thanks{Corresponding author, {\tt broggini@pd.infn.it}.}
 \affiliation{Istituto Nazionale di Fisica Nucleare (INFN), Sezione di Padova, via Marzolo 8, 35131 Padova, Italy}
\author{P.~Corvisiero}\affiliation{Universit\`a di Genova and INFN Sezione di Genova, Genova, Italy}
\author{Z.~Elekes}\affiliation{Institute of Nuclear Research (ATOMKI), Debrecen, Hungary}
\author{Zs.~F\"ul\"op}\affiliation{Institute of Nuclear Research (ATOMKI), Debrecen, Hungary}
\author{G.~Gervino}\affiliation{Dipartimento di Fisica Sperimentale, Universit\`a di Torino and INFN Sezione di Torino, Torino, Italy}
\author{A.~Guglielmetti}\affiliation{Istituto di Fisica Generale Applicata, Universit\`a di Milano and INFN Sezione di Milano, Italy}
\author{C.~Gustavino}\affiliation{INFN, Laboratori Nazionali del Gran Sasso (LNGS), Assergi (AQ), Italy}
\author{G.~Imbriani}\affiliation{Dipartimento di Scienze Fisiche, Universit\`a di Napoli ''Federico II'', and INFN Sezione di Napoli, Napoli, Italy}
\author{M.~Junker}\affiliation{INFN, Laboratori Nazionali del Gran Sasso (LNGS), Assergi (AQ), Italy}
\author{M.~Laubenstein}\affiliation{INFN, Laboratori Nazionali del Gran Sasso (LNGS), Assergi (AQ), Italy}
\author{A.~Lemut}\affiliation{Universit\`a di Genova and INFN Sezione di Genova, Genova, Italy}
\author{B.~Limata}\affiliation{Dipartimento di Scienze Fisiche, Universit\`a di Napoli ''Federico II'', and INFN Sezione di Napoli, Napoli, Italy}
\author{V.~Lozza}\affiliation{Istituto Nazionale di Fisica Nucleare (INFN), Sezione di Padova, via Marzolo 8, 35131 Padova, Italy}
\author{M.~Marta}\affiliation{Istituto di Fisica Generale Applicata, Universit\`a di Milano and INFN Sezione di Milano, Italy}
\author{R.~Menegazzo}\affiliation{Istituto Nazionale di Fisica Nucleare (INFN), Sezione di Padova, via Marzolo 8, 35131 Padova, Italy}
\author{P.~Prati}\affiliation{Universit\`a di Genova and INFN Sezione di Genova, Genova, Italy}
\author{V.~Roca}\affiliation{Dipartimento di Scienze Fisiche, Universit\`a di Napoli ''Federico II'', and INFN Sezione di Napoli, Napoli, Italy}
\author{C.~Rolfs}\affiliation{Institut f\"ur Experimentalphysik III, Ruhr-Universit\"at Bochum, Bochum, Germany}
\author{C.~Rossi Alvarez}\affiliation{Istituto Nazionale di Fisica Nucleare (INFN), Sezione di Padova, via Marzolo 8, 35131 Padova, Italy}
\author{E.~Somorjai}\affiliation{Institute of Nuclear Research (ATOMKI), Debrecen, Hungary}
\author{O.~Straniero}\affiliation{Osservatorio Astronomico di Collurania, Teramo, and INFN Sezione di Napoli, Napoli, Italy}
\author{F.~Strieder}\affiliation{Institut f\"ur Experimentalphysik III, Ruhr-Universit\"at Bochum, Bochum, Germany}
\author{F.~Terrasi}\affiliation{Seconda Universit\`a di Napoli, Caserta, and INFN Sezione di Napoli, Napoli, Italy}
\author{H.P.~Trautvetter}\affiliation{Institut f\"ur Experimentalphysik III, Ruhr-Universit\"at Bochum, Bochum, Germany}

\collaboration{The LUNA Collaboration}\noaffiliation

\date{July 5, 2006}
%\preprint{Version accepted by Physical Review Letters on August 17, 2006}

\begin{abstract}
The nuclear physics input from the $^3$He($\alpha,\gamma$)$^7$Be
cross section is a major uncertainty in the fluxes of $^7$Be and
$^8$B neutrinos from the Sun predicted by solar models and in the
$^7$Li abundance obtained in big-bang nucleosynthesis calculations.
The present work reports on a new precision experiment using the
activation technique at energies directly relevant to big-bang nucleosynthesis.
Previously such low energies had been reached experimentally only by
the prompt-$\gamma$ technique and with inferior precision.
Using a windowless gas target, high beam intensity and low background
$\gamma$-counting facilities, the $^3$He($\alpha,\gamma$)$^7$Be cross
section has been determined at 127, 148 and 169\,keV center-of-mass energy
with a total uncertainty of 4\,\%. The sources of systematic uncertainty
are discussed in detail.
The present data can be used in big-bang nucleosynthesis
calculations and to constrain the extrapolation of the
$^3$He($\alpha,\gamma$)$^7$Be astrophysical S-factor to solar
energies.
\end{abstract}

\pacs{25.55.-e, 26.20.+f, 26.35.+c, 26.65.+t}

\maketitle

The $^3$He($\alpha,\gamma$)$^7$Be reaction is a critical link in the $^7$Be
and $^8$B branches of the proton--proton (p--p) chain of solar hydrogen
burning \cite{BS05}.
At low energies its cross section $\sigma(E)$ ($E$ denotes the center
of mass energy, $E_\alpha$ the $^4$He beam energy in the laboratory system)
can be parameterized by the astrophysical S-factor $S(E)$ defined as
$$ S(E) = \sigma(E) \cdot E \exp(2\pi\eta(E)) $$
with $\eta(E) \propto E^{-0.5}$ \cite{Rolfs88}. The 9.4\,\%
uncertainty \cite{Adelberger98-RMP} in the S-factor extrapolation to
the solar Gamow energy (23\,keV) contributes 8\,\% to the uncertainty in the
predicted fluxes of solar neutrinos from the decays of $^7$Be and
$^8$B \cite{Bahcall04-PRL}.
The interior of the Sun, in turn, can be studied
\cite{Bahcall04-PRL,Fiorentini03-arxiv} by comparing this prediction with
the data from neutrino detectors \cite{Ahmed04-PRL,SuperKamiokande06-PRD},
which determine the $^8$B neutrino flux with a total uncertainty as low as 3.5\,\%
\cite{SuperKamiokande06-PRD}.

Furthermore, the production of $^7$Li in big-bang nucleosynthesis (BBN) is
highly sensitive to the $^3$He($\alpha,\gamma$)$^7$Be cross section in the
energy range $E$ $\approx$ 160--380\,keV \cite{Burles99-PRL}, with an
adopted uncertainty of 8\,\% \cite{Descouvemont04-ADNDT}. Based on the
baryon-to-photon ratio from observed anisotropies in the cosmic microwave
background \cite{Spergel03-ApJSS}, network calculations predict primordial
$^7$Li abundances \cite{Coc04-ApJ} that are significantly higher than
observations \cite{Ryan00-ApJL,Bonifacio02-AA}.
A lower $^3$He($\alpha$,$\gamma$)$^7$Be cross section at relevant energies
may explain part of this discrepancy.

The $^3$He($\alpha$,$\gamma$)$^7$Be ($Q$-value: 1.586\,MeV) reaction leads
to the emission of prompt $\gamma$-rays, and the final $^7$Be nucleus decays
with a half-life of 53.22\,$\pm$\,0.06\,days, emitting a 478\,keV $\gamma$-ray
in 10.44\,$\pm$\,0.04\,\% of the cases \cite{Tilley02-NPA}. The cross section
can be measured by detecting either the induced $^7$Be activity (activation
method) or the prompt $\gamma$-rays from the reaction (prompt-$\gamma$
method). Previous activation studies
\cite{Osborne82-PRL,Robertson83-PRC,Volk83-ZPA,NaraSingh04-PRL} cover the
energy range $E$ = 420--2000\,keV. Prompt $\gamma$-ray measurements
\cite{Holmgren59-PR,Parker63-PR,Nagatani69-NPA,Kraewinkel82-ZPA,Osborne82-PRL,Alexander84-NPA,Hilgemeier88-ZPA}
cover $E$ = 107--2500\,keV, although with limited precision at low energies.

%%%%%%%%%%%%%%%%%%
\begin{figure*}[tb]
 \centering
  \includegraphics[angle=270,width=1.0\textwidth]{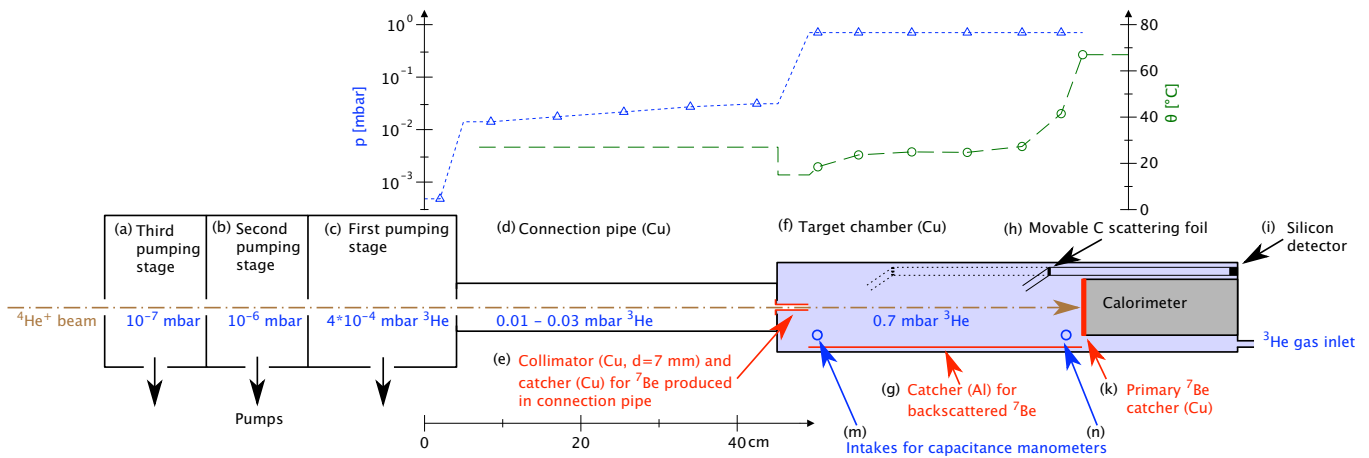}
 \caption{\label{fig:setup} Schematic view of the target chamber used for the irradiations. Above: pressure ($p$, triangles) and temperature ($\theta$, circles) values measured without ion beam and interpolated profile between the data points (lines). See text for details.
  }
\end{figure*}

The global shape of the S-factor curve is well reproduced by
theoretical calculations \cite{Kajino86-NPA,Csoto00-FBS}. However,
the slope has been questioned \cite{Csoto00-FBS} for $E$ $\leq$
300\,keV, where there are no high-precision data. Furthermore, a
global analysis \cite{Adelberger98-RMP} indicates that S-factor data
obtained with the activation method are systematically higher than
the prompt-$\gamma$ results. A recent activation study
\cite{NaraSingh04-PRL} reduces this discrepancy to 9\,\% for the
extrapolated $S(0)$ \cite{Adelberger98-RMP}, still not at the
precision level of the $^8$B neutrino data
\cite{SuperKamiokande06-PRD}.
Precise $^3$He($\alpha$,$\gamma$)$^7$Be measurements at low energies have been recommended to study the solar interior
\cite{Bahcall04-PRL,Fiorentini03-arxiv,Bahcall05-astroph},
to sharpen big-bang $^7$Li abundance predictions
\cite{Burles99-PRL,Serpico04-JCAP}, and to investigate the low-energy slope of the S-factor curve \cite{Csoto00-FBS}. The aim of the present work is to provide high precision activation data at energies directly relevant to big-bang nucleosynthesis and low enough to effectively constrain the extrapolation to solar energies.

The Laboratory for Underground Nuclear Astrophysics (LUNA) \cite{Greife94-NIMA} in Italy's Gran Sasso underground laboratory (LNGS) has been designed for measuring low nuclear cross sections for astrophysical purposes \cite{Bonetti99-PRL,Casella02-NPA,Formicola04-PLB,Bemmerer05-EPJA,Imbriani05-EPJA,Lemut06-PLB}. The irradiations have been carried out at the 400\,kV LUNA2 accelerator \cite{Formicola03-NIMA} at energies $E_\alpha$ = 300, 350 and 400\,keV, with a typical current of 200~$\mu$A $^4$He$^+$. The beam energy is obtained from a precision resistor chain and has 5\,eV/h long-term stability \cite{Formicola03-NIMA}.
The $^3$He($\alpha,\gamma$)$^7$Be reaction takes place in a differentially pumped windowless gas target (Fig.~\ref{fig:setup}, similar to the one described previously \cite{Casella02-NIMA}) filled with enriched $^3$He gas (isotopic purity $>$99.95\,\%, pressure 0.7\,mbar, target thickness 9--10\,keV). The exhaust from the first and second pumping stages is cleaned in a getter-based gas purifier and recirculated into the target. The ion beam from the accelerator passes three pumping stages (Fig.~\ref{fig:setup}\,a-c),\linebreak a connection pipe (d), enters the target chamber (f) through an aperture of 7\,mm diameter (e) and is finally stopped on a detachable oxygen free high conductivity (OFHC) copper disk (k) of 70\,mm diameter that serves as the primary catcher for the produced $^7$Be and as the hot side of a calorimeter with constant temperature gradient \cite{Casella02-NIMA}. A precision of 1.5\,\% for the beam intensity is obtained from the difference between the calorimeter power values with and without incident ion beam, taking into account the calculated energy loss in the target gas \cite{SRIM03-26} and using a calibration curve determined by measuring the electrical charge in the same setup without gas, applying a proper secondary electron suppression voltage.
The effective target thickness depends on the pressure (monitored during the irradiations with two capacitance manometers, Fig.~\ref{fig:setup}\,m-n), the pressure and temperature profile (measured without ion beam, resulting density uncertainty 0.6\,\%), the thinning of the target gas through the beam heating effect \cite{Goerres80-NIM} and the fraction of gases other than $^3$He. In order to study the latter two effects, a 100\,$\mu$m thick silicon detector (Fig.~\ref{fig:setup}\,i) detects projectiles that have been elastically scattered first in the target gas and subsequently in a movable 15\,$\mu$g/cm$^2$ carbon foil (h). The beam heating effect has been investigated in a wide beam energy and intensity range, and a correction of 4.9$\pm$1.3\,\%, 5.4$\pm$1.3\,\% and 5.7$\pm$1.3\,\% was found for the irradiations at $E_\alpha$ = 300, 350 and 400\,keV, respectively. The amount of contaminant gases (mainly nitrogen) is monitored with the silicon detector during the irradiations, kept below 1.0$\pm$0.1\,\% and corrected for in the analysis. Further details of the elastic scattering measurements are described elsewhere \cite{Marta05-Master}.

The catchers are irradiated with charges of 60--220\,C, accumulating $^7$Be activities of 0.2--0.5\,Bq.
The effective center of mass energy $E^{\rm eff}$ is calculated assuming a constant S-factor over the target length \cite{Rolfs88}. The uncertainties of 0.3\,keV in $E_\alpha$ \cite{Formicola03-NIMA} and of 4.4\,\% in the energy loss \cite{SRIM03-26} result in an S-factor uncertainty of 0.5--0.8\,\%.
Calculations for the straggling of the $^4$He beam and of the produced $^7$Be nuclei in the $^3$He gas and for the emission cone of $^7$Be (opening angle 1.8--2.1\,$^\circ$) show that 99.8\,\% of the $^7$Be produced inside the target chamber, including the 7\,mm collimator, reaches the primary catcher.

\begin{figure}
 \centering
\includegraphics[angle=0,width=\columnwidth]{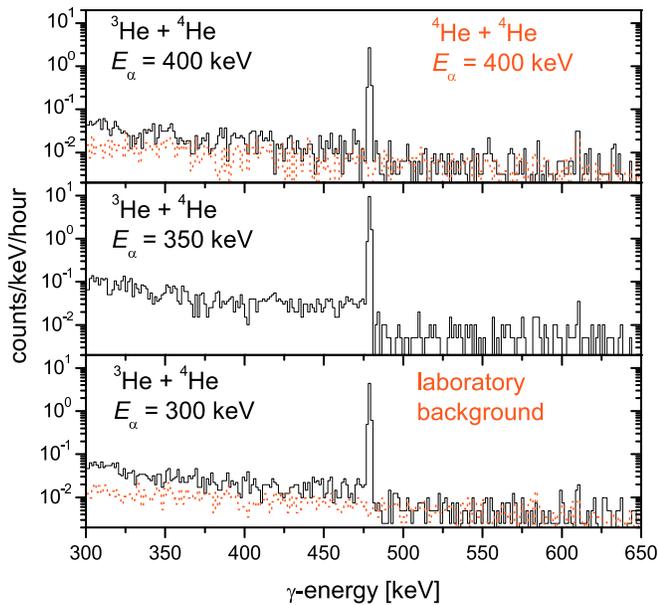}
 \caption{\label{fig:spectrum}
Offline $\gamma$-counting spectra, detector LNGS1. Solid black line: $^3$He gas bombarded at $E_\alpha$ = 400, 350, 300\,keV (top to down), respectively. Dotted red line, top panel: $^4$He gas bombarded at $E_\alpha$ = 400\,keV. Dotted red line, bottom panel: laboratory background.}
\end{figure}

After the irradiation, the catcher is dismounted and counted in close geometry subsequently with two 120\,\% relative efficiency HPGe detectors called LNGS1  (Fig.~\ref{fig:spectrum}) and LNGS2, both properly shielded with copper and lead, in the LNGS underground counting facility \cite{Arpesella96-Apradiso}. Detector LNGS1 is additionally equipped with an antiradon box, and
its laboratory background is two orders of magnitude lower than with equivalent shielding overground \cite{Arpesella96-Apradiso}.
In order to obtain the photopeak counting efficiencies, three homogeneous $^7$Be sources of 200--800\,Bq activity and 8\,mm active diameter were prepared with the $^7$Li(p,n)$^7$Be reaction at ATOMKI. Their activity was determined with two HPGe detectors (each efficiency based on an independent set of commercial $\gamma$-ray sources) at ATOMKI and with one HPGe detector, called LNGS3  (efficiency based on a third set of commercial sources), at LNGS, giving consistent results and a final activity uncertainty of 1.8\,\%. The three
% ATOMKI
$^7$Be sources were then used to calibrate detectors LNGS1 and LNGS2 in the same geometry as the activated samples. The $^7$Be distribution in the catchers has been calculated from the $^7$Be emission angle and straggling, and GEANT4 \cite{Agostinelli03-NIMA} simulations gave 0.8$\pm$0.4\,\% to 1.0$\pm$0.4\,\% correction for the $\gamma$-ray efficiency because of the tail of the distribution at high radii.

\begin{table}
\caption{\label{tab:uncert} Systematic uncertainties in the $^3$He($\alpha,\gamma$)$^7$Be astrophysical S-factor, neglecting contributions below 0.2\,\%.}

\begin{ruledtabular}
\setlength{\extrarowheight}{0.1cm}
\begin{tabular}{lc}
\multicolumn{1}{c}{Source} & Uncertainty\\
\hline
$^7$Be counting efficiency & 1.8\,\%\\
Beam intensity & 1.5\,\%\\
Beam heating effect  & 1.3\,\%\\
Target pressure and temperature without beam & 0.6\,\%\\
$^7$Be backscattering & 0.5\,\%\\
Incomplete $^7$Be collection & 0.4\,\%\\
$^7$Be distribution in catcher & 0.4\,\%\\
478\,keV $\gamma$-ray branching \cite{Tilley02-NPA} & 0.4\,\%\\
Effective energy & 0.5--0.8\,\%\\
\hline
Total: & 2.9--3.0\,\%\\
\end{tabular}
\end{ruledtabular}
\end{table}

\begin{figure}[ht!]
 \centering
  \includegraphics[angle=270,width=\columnwidth]{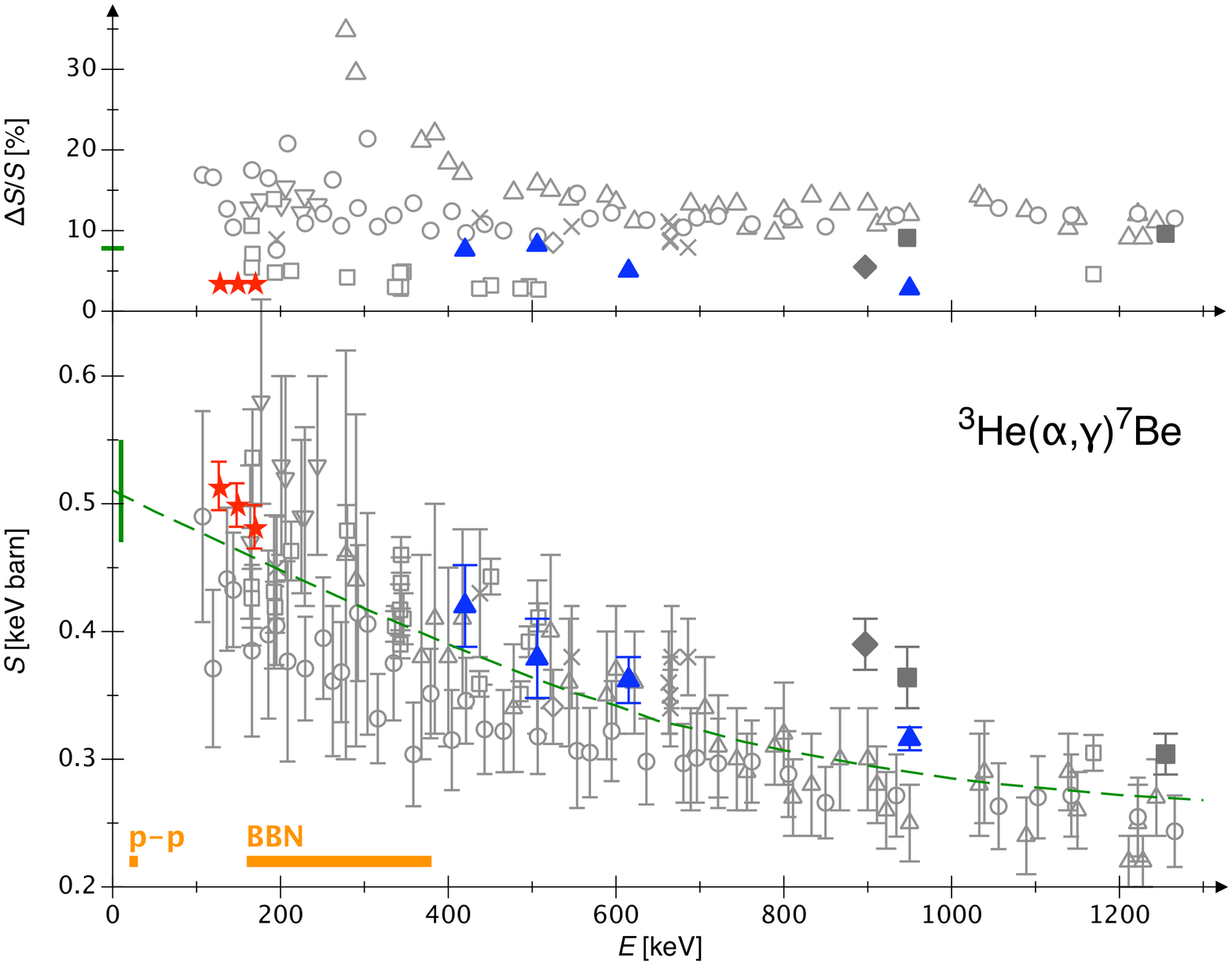}
 \caption{\label{fig:sfactor} Lower panel: astrophysical S-factor for $^3$He($\alpha,\gamma$)$^7$Be.
Activation data: filled squares \cite{Osborne82-PRL}, filled diamonds \cite{Robertson83-PRC}, filled triangles \cite{NaraSingh04-PRL}, stars (present work).
Prompt-$\gamma$ data: triangles \cite{Parker63-PR}, inverted triangles \cite{Nagatani69-NPA}, circles \cite{Kraewinkel82-ZPA} (renormalized by a factor 1.4 \cite{Hilgemeier88-ZPA}), squares \cite{Osborne82-PRL}, diamonds \cite{Alexander84-NPA}, crosses \cite{Hilgemeier88-ZPA}.
Dashed line: previously adopted R-matrix fit \cite{Descouvemont04-ADNDT}. Horizontal bars: energies relevant for p--p chain and for BBN.
--- Upper panel: uncertainties (systematic and statistical combined in quadrature) of the data and of the R-matrix $S(0)$ \cite{Descouvemont04-ADNDT}.
 }
\end{figure}

In order to investigate parasitic production of $^7$Be through,
e.g., the $^6$Li(d,n)$^7$Be and $^{10}$B(p,$\alpha$)$^7$Be reactions
induced by possible traces of $^2$DH$_2^+$ in the $^4$He$^+$ beam,
the enriched $^3$He target gas was replaced with 0.7\,mbar $^{4}$He,
and a catcher was bombarded at the highest available energy of
$E_\alpha$ = 400\,keV. Despite the high applied dose of 104\,C, in
16 days counting time no $^7$Be has been detected
(Fig.~\ref{fig:spectrum}, top panel), establishing a 2$\sigma$ upper
limit of 0.1\% for parasitic $^7$Be.

Furthermore, $^7$Be losses by backscattering from the primary
catcher and by incomplete collection were studied experimentally at
$E_\alpha$ = 400\,keV and with Monte Carlo simulations at 300, 350
and 400\,keV. For the backscattering study, parts of the inner
surface of the chamber were covered by aluminum foil functioning as
secondary catcher (Fig.~\ref{fig:setup}\,g). It was found that
1.3\,$\pm$\,0.5\,\% of the created $^7$Be is lost due to
backscattering, consistent with 1.5\,\% obtained in a GEANT4
\cite{Agostinelli03-NIMA} simulation using a SRIM-like multiple
scattering process \cite{Mendenhall05-NIMB}. At lower energies, the
simulation result was used as backscattering correction (up to
2.2\,\%, adopted uncertainty 0.5\,\%).

Incomplete $^7$Be collection occurs since 3.5\,\% of the total
$^3$He target thickness are in the connecting pipe, and a part of
the $^7$Be created there does not reach the primary catcher but is
instead implanted into the 7\,mm collimator
(Fig.~\ref{fig:setup}\,e). At $E_\alpha$ = 400\,keV, a modified
collimator functioning as secondary catcher was used, and a
2.6\,$\pm$\,0.4\,\% effect was observed, consistent with a
simulation (2.1$\pm$0.4\,\%). For $E_\alpha$ = 300 and 350\,keV,
incomplete $^7$Be collection was corrected for based on the
simulation (up to 2.3\,\% correction, adopted uncertainty 0.4\,\%).

Sputtering losses of $^7$Be by the $^4$He beam were simulated
\cite{SRIM03-26}, showing that for the present beam energies
sputtering is 10$^{4}$ times less likely than transporting the
$^7$Be even deeper into the catcher, so it has been neglected.

The systematic uncertainties are summarized in
Table~\ref{tab:uncert}, giving a total value of 3\,\%. For the
present low energies an electron screening enhancement factor $f$
\cite{Assenbaum87-ZPA} of up to 1.012 has been calculated in the
adiabatic limit, but not corrected for (Table~\ref{tab:results}).

The present data (Table~\ref{tab:results}, lower panel of
Fig.~\ref{fig:sfactor}) are the first activation results at energies
directly relevant to big-bang $^7$Li production. Their uncertainty
of 4\,\% (systematic and statistical combined in quadrature) is
comparable to or lower than previous activation studies at high
energy and lower than prompt-$\gamma$ studies at comparable energy
(upper panel of Fig.~\ref{fig:sfactor}).

%%%%%%%%%%%%%%%%%%%%%%%%%%%%%%%%
\begin{table}
\caption{\label{tab:results} Cross section and S-factor results,
relative uncertainties, and electron screening
\cite{Assenbaum87-ZPA} enhancement factors $f$.}
\begin{ruledtabular}
\setlength{\extrarowheight}{0.1cm}
\begin{tabular}{ccccccc}
$E^{\rm eff}$ & $\sigma(E^{\rm eff})$ & $S(E^{\rm eff})$ & \multicolumn{2}{c}{$\Delta$$S$/$S$} & $f$ \\
\ [keV] & [10$^{-9}$ barn] & [keV barn] & stat. & syst. & \\
\hline
126.5 & 1.87 & 0.514 & 2.0\,\% & 3.0\,\% & 1.012 \\
147.7 & 4.61 & 0.499 & 1.7\,\% & 2.9\,\% & 1.009 \\
168.9 & 9.35 & 0.482 & 2.0\,\% & 2.9\,\% & 1.008 \\
\end{tabular}
\end{ruledtabular}
\end{table}

To give an estimate for the low-energy implications, rescaling the
most recent R-matrix fit \cite{Descouvemont04-ADNDT} to the present
data results in $S(0)$ = 0.547$\pm$0.017\,keV\,barn, consistent
with, but more precise than, Ref. \cite{NaraSingh04-PRL}. All
activation data combined (Refs.
\cite{Osborne82-PRL,Robertson83-PRC,Volk83-ZPA,NaraSingh04-PRL} and
the present work) give $S(0)$ = 0.550$\pm$0.012\,keV\,barn, higher
than the weighted average of all previous prompt-$\gamma$ studies,
$S(0)$ =  0.507$\pm$0.016\,keV\,barn \cite{Adelberger98-RMP}.
Prompt-$\gamma$ experiments with precision comparable to the 4\,\%
reached in the present activation work are now called for in order
to verify the normalization of the prompt-$\gamma$ data.

\begin{acknowledgments}
This work was supported by INFN and in part by: TARI RII-CT-2004-506222, OTKA T42733 and T49245, and BMBF (05CL1PC1-1).
\end{acknowledgments}

%%%%%%%%%%%%%%%%%%%%%%%%%%%%%%%%
% \bibliography{actibib}

\end{document}